\documentclass[prb,twocolumn,showpacs,preprintnumbers,amsmath,amssymb]{revtex4-1}
\usepackage{amssymb}
\usepackage{bm}
\usepackage{amsmath}
\usepackage[dvips]{graphicx}
\usepackage{color}
\usepackage{comment}
\usepackage{multirow}
\usepackage{subfigure}

\begin{document}
\title{Chern insulating state in laterally patterned semiconductor heterostructures}

\author{Tommy Li$^1$,  Oleg P. Sushkov$^1$}
\affiliation{$^1$School of Physics, University of New South Wales, Sydney 2052, Australia}
\begin{abstract}
Hexagonally patterned two-dimensional $p$-type semiconductor systems are quantum simulators  of graphene with strong and highly tunable spin-orbit interactions. We show that application of purely in-plane magnetic fields, in combination with the crystallographic anisotropy present in low-symmetry semiconductor interfaces,  allows Chern insulating phases to emerge from an originally topologically insulating state after a quantum phase transition. These phases are characterized by pairs of co-propagating edge modes and Hall conductivities $\sigma_{xy} = +\frac{2 e^2}{h}, -\frac{2 e^2}{h}$ in the absence of Landau levels or cyclotron motion. With current lithographic technology, the Chern insulating transitions are predicted to occur in GaAs heterostructures at magnetic fields of $\sim 5\text{T}$.
\end{abstract}

\maketitle

\begin{figure*}[t]
	\begin{tabular}{cc}
		\includegraphics[width = 0.36\textwidth]{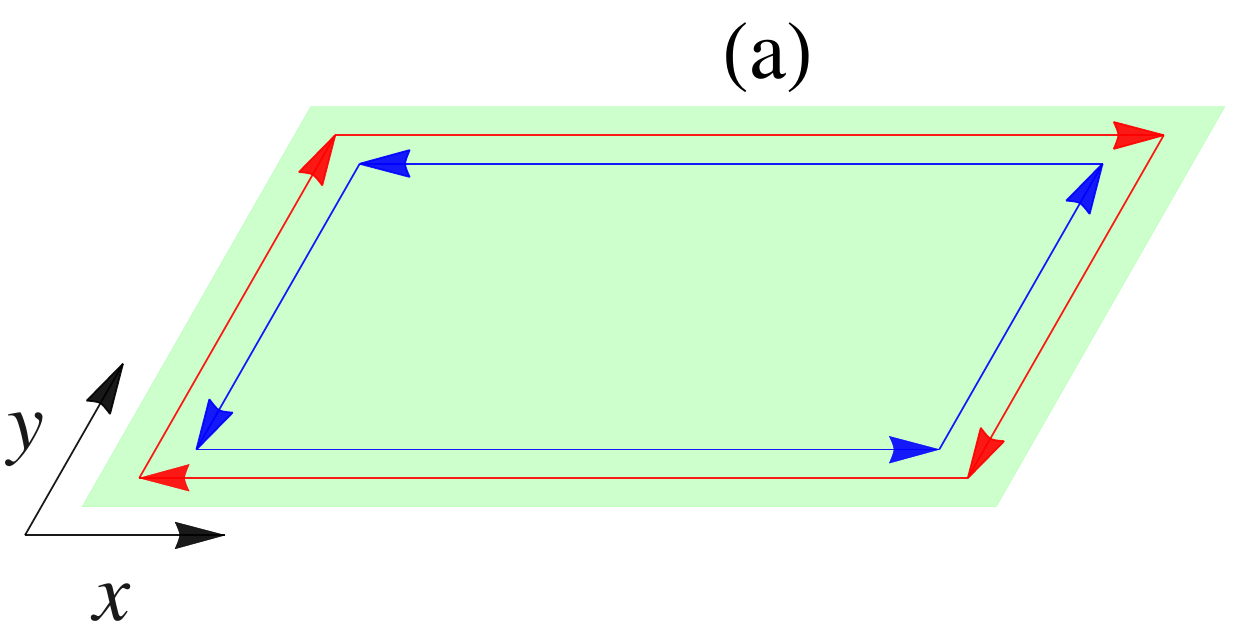}
		& \includegraphics[width = 0.35\textwidth]{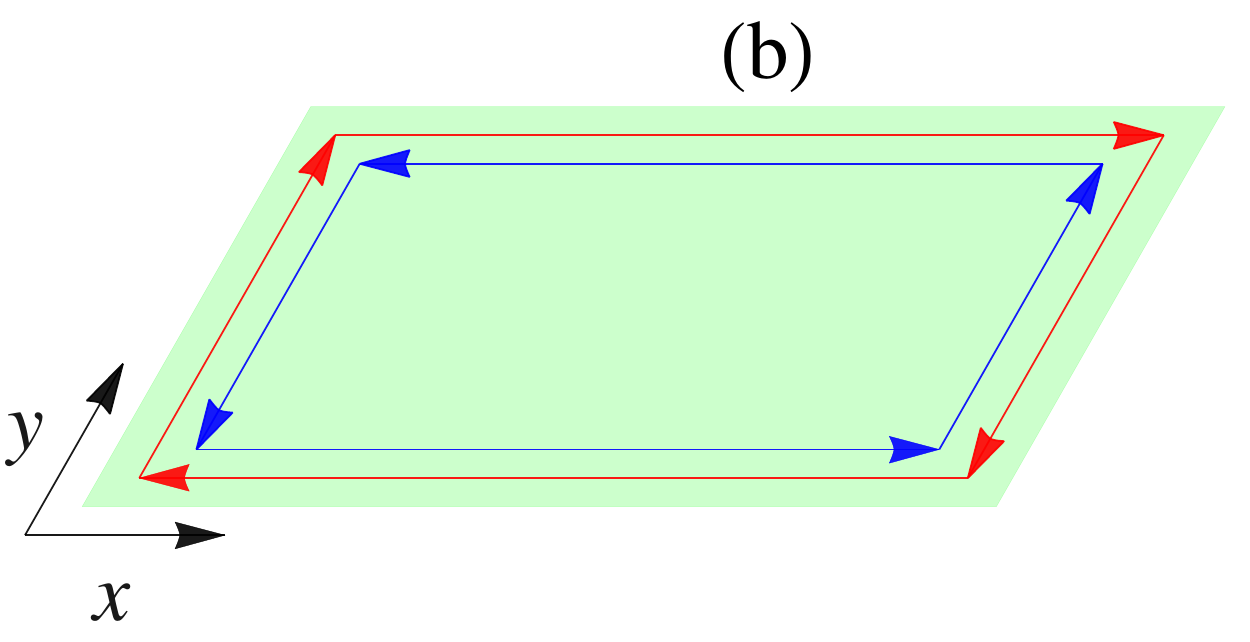} \\
		\includegraphics[width = 0.50\textwidth]{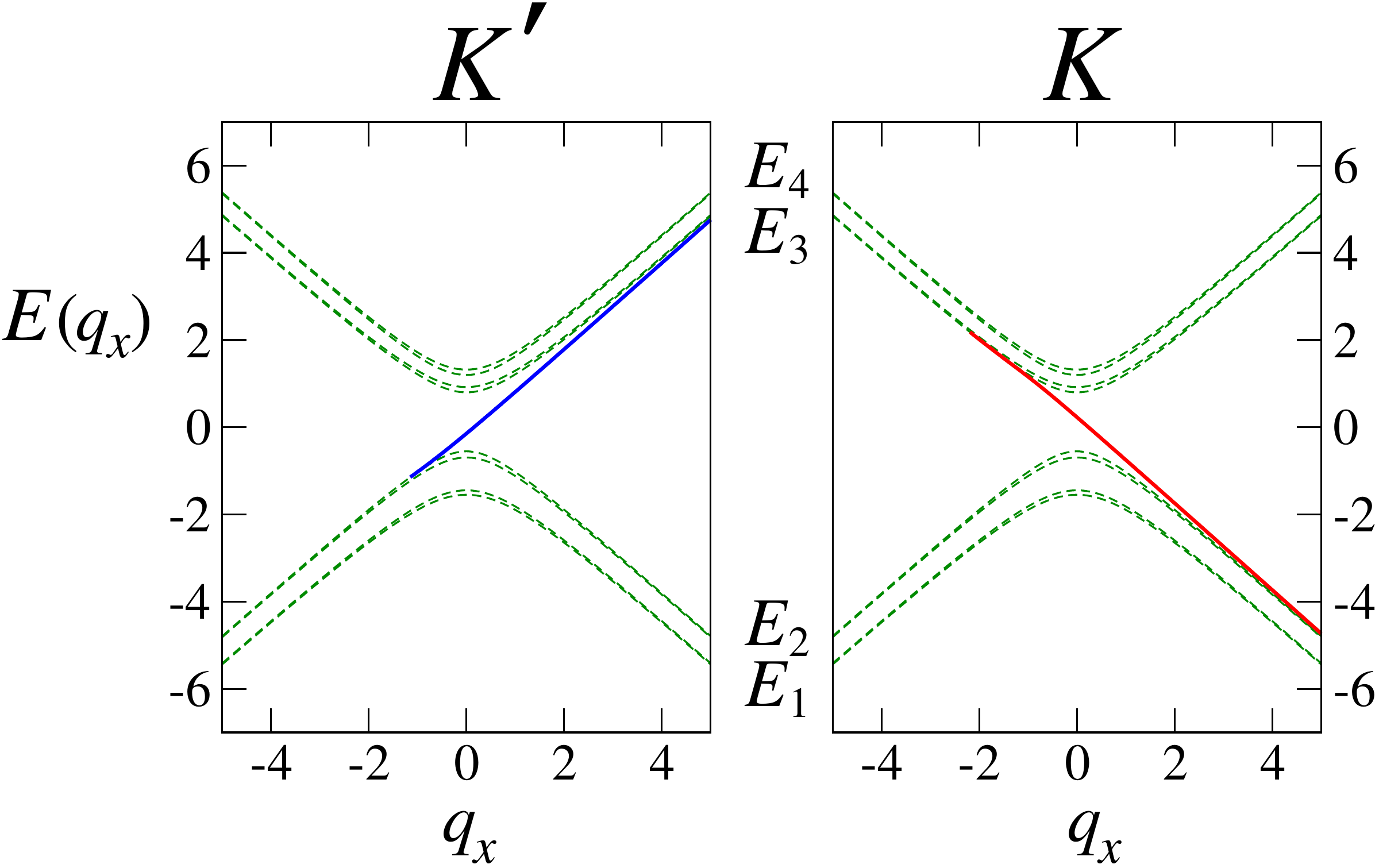}
		&
		 \includegraphics[width = 0.495\textwidth]{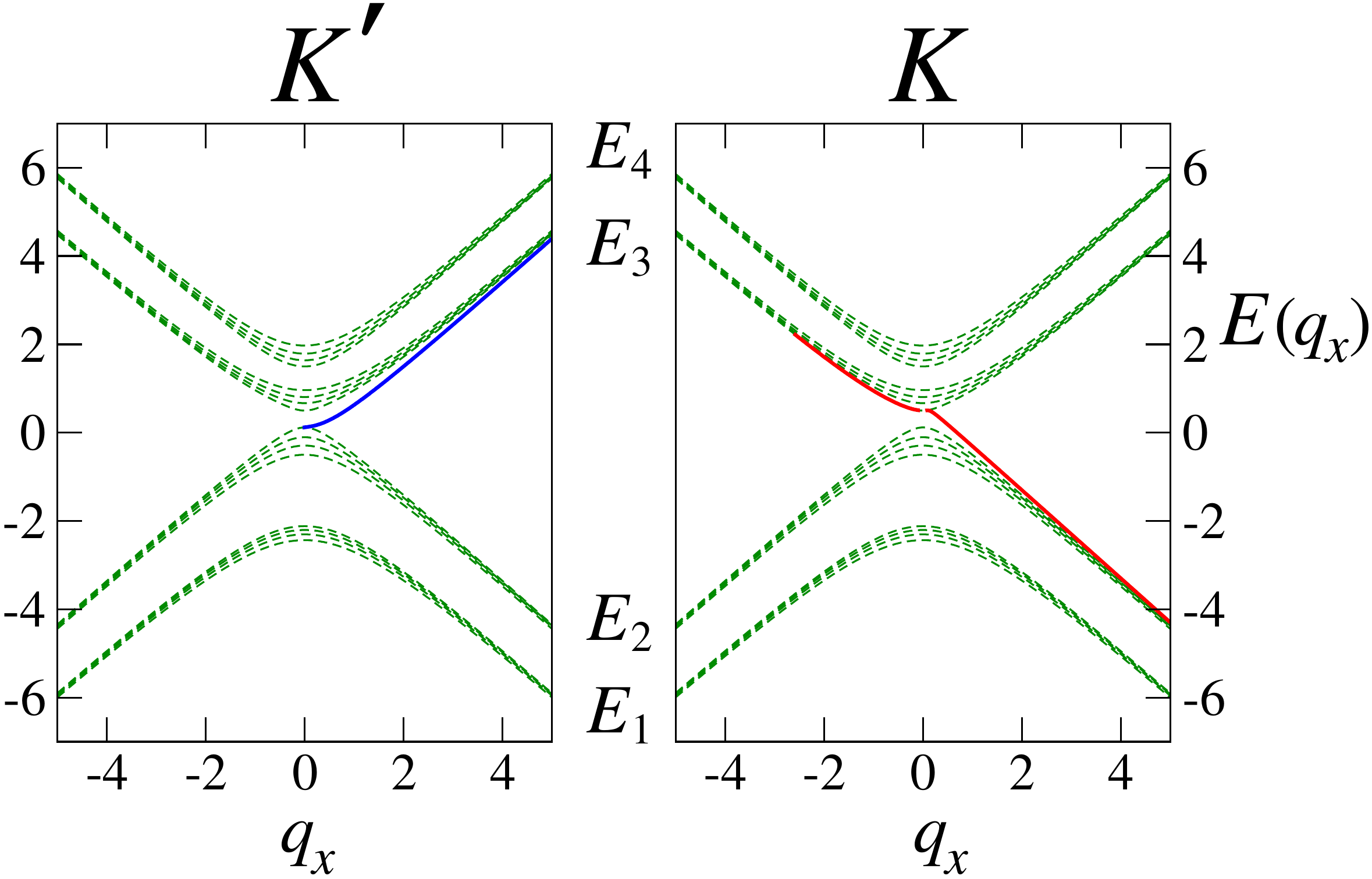}  \\
		\includegraphics[width = 0.36 \textwidth]{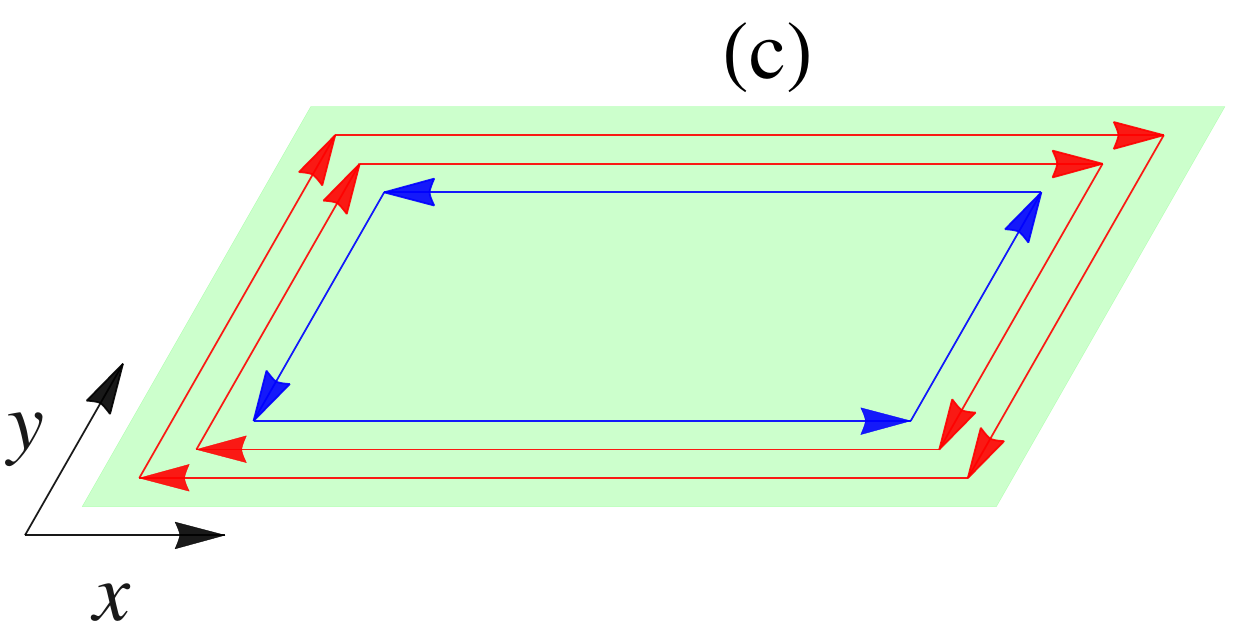}
		& \includegraphics[width = 0.35\textwidth]{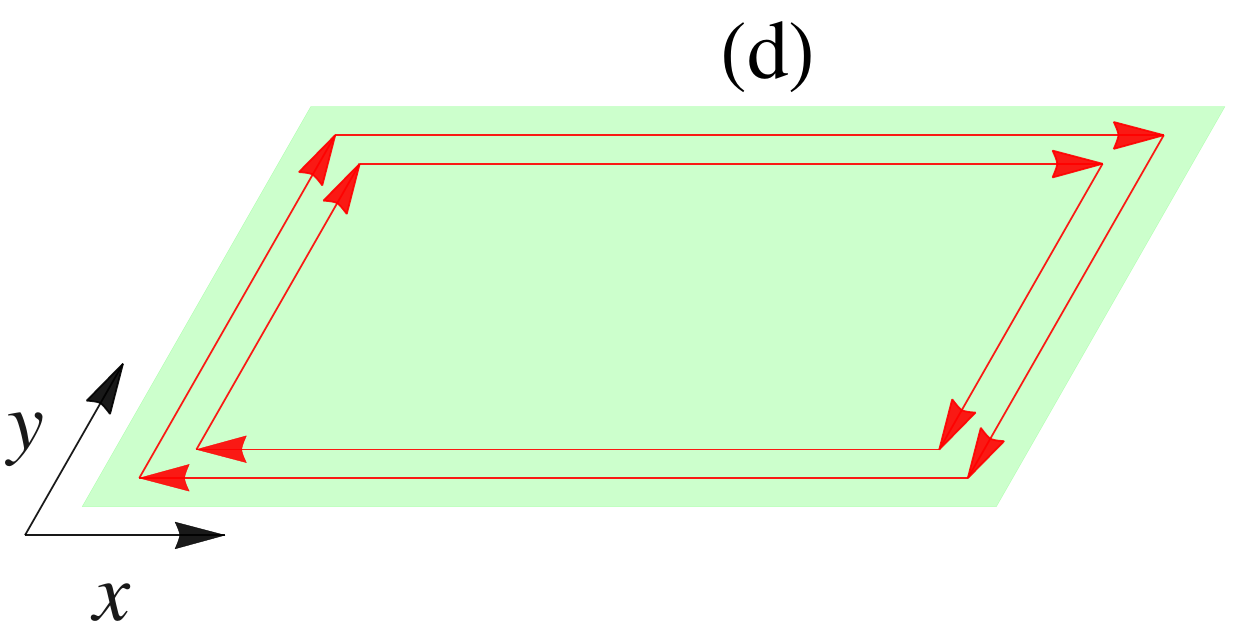} \\
		\includegraphics[width = 0.50\textwidth]{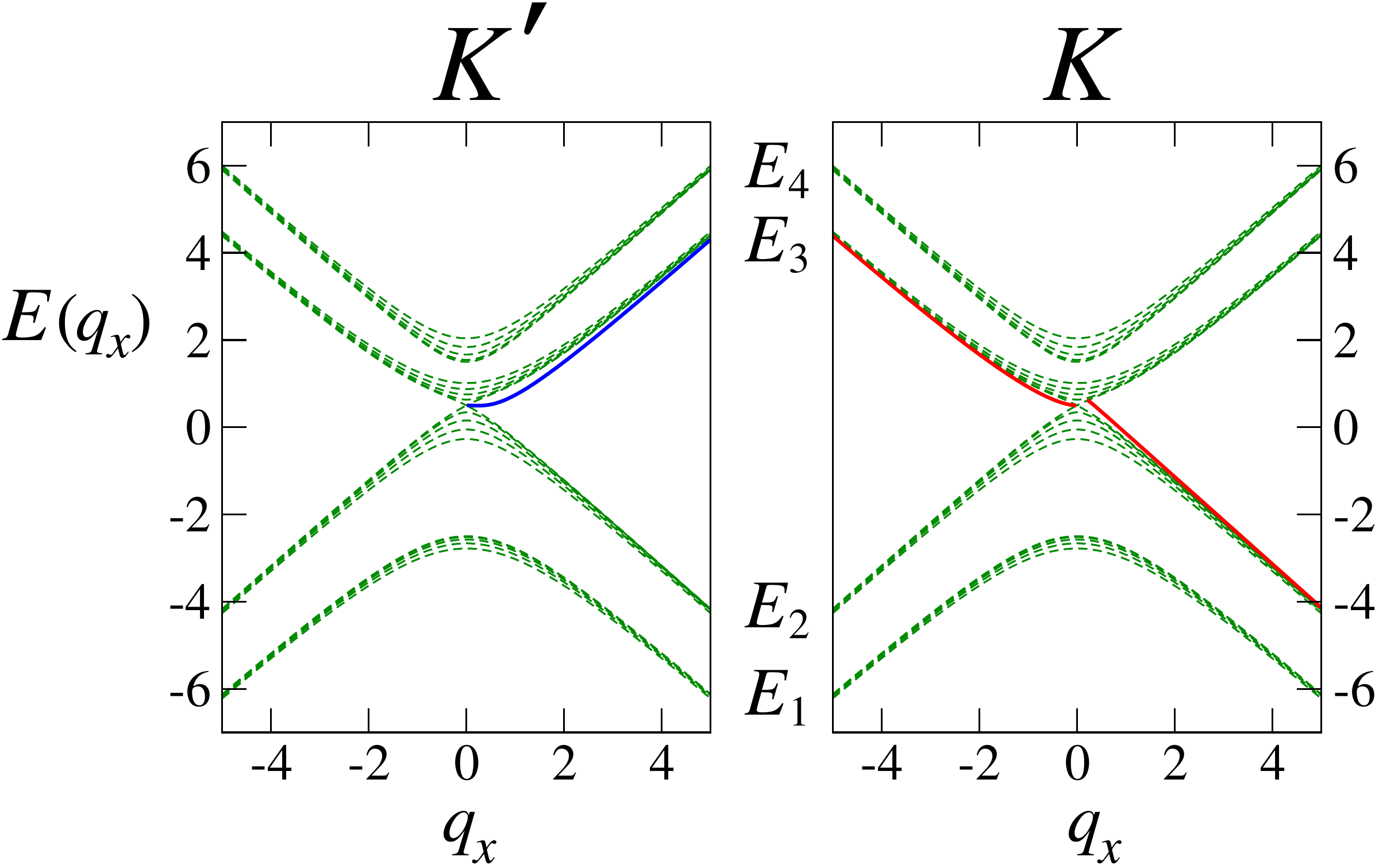}
		& \includegraphics[width = 0.495\textwidth]{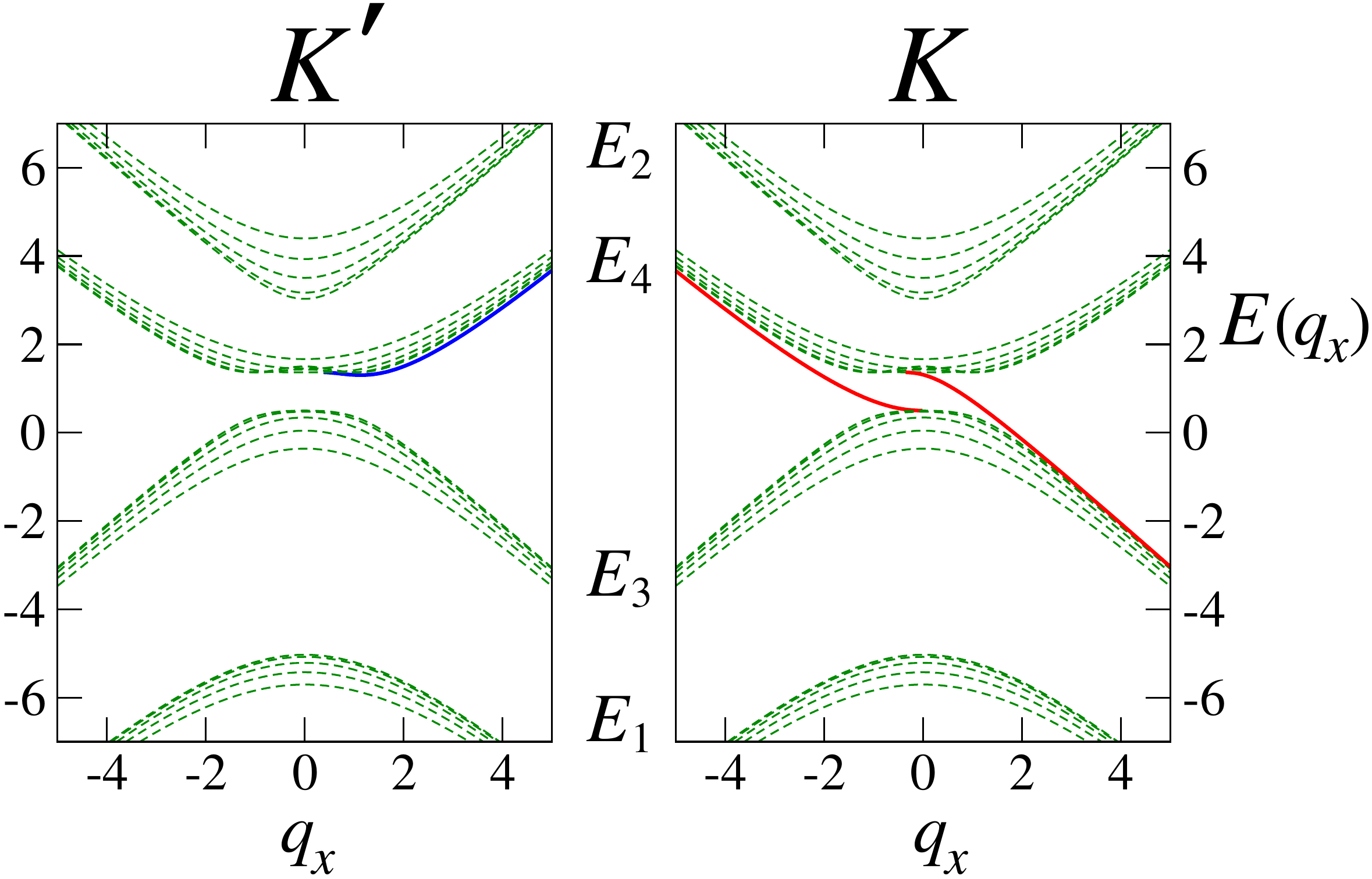}
	\end{tabular}
	\caption{Edge modes (solid lines) along a hard-wall boundary along the $x$-direction, calculated for parameters: (a) $\beta = 0.2\eta, \gamma = 0.2\eta$, (b) $\beta = 0.5 \eta, \gamma = 0.5\eta$, (c) $\beta = \sqrt{2} \eta, \gamma = \frac{ \eta}{2}$, (d)  $\beta = 2\eta, \gamma = 0.5\eta$. Edge modes consisting of Bloch wavefunctions at the $K$ ($K'$) valley are indicated in red (blue). The dashed lines show the bulk states for various values of $q_y$. Panels (a),(b) correspond to the insulating state with Chern number $N = 0$, $\beta^2 + \eta \gamma < \eta^2$, (c) corresponds to a critical gapless state $\beta = \sqrt{ \eta^2 - \eta \gamma}$, and (d) corresponds to the Chern insulating state, $\beta^2 + \eta \gamma> \eta^2$. The axes correspond to units in which $\eta = 1, v = 1$ (so  the vertical axis shows $\frac{E(q_x)}{ \eta}$ and the horizontal axis shows $\frac{v q_x}{\eta}$). The scale of both axes is the same in (a) and (b).} 
	\end{figure*}

The versatility of the spin-orbit interaction in  semiconductor heterostructures has made them the subject of numerous recent theoretical proposals to access novel states of matter\cite{bilayerexciton,QSHStrain,QSH,QSHHgTe,QSHInAs,TopSup,TopSup2,Majorana}. These theoretical efforts are supported by recent observations of the Quantum Spin Hall Insulator phase (first predicted to exist in graphene\cite{KaneMele}) in HgTe\cite{QSHobs} and InAs\cite{QSHobsInAs} quantum wells, proving them to be promising candidates for the realization of topologically protected room-temperature spin transport. While spin-polarized edge transport associated with the topological insulator phase in InAs quantum wells has been the subject of several investigations following its initial discovery\cite{QSHobsInAs2,InAsEdge,InAsLutt}, measurement of lateral spin accumulation, a fundamental signature of the Quantum Spin Hall  phase,\cite{KaneMele} remains an open challenge, since measurement of the spin polarization of edge states requires coupling to an external magnetic field (e.g. coupling to ferromagnetic leads\cite{QSHprobe}), which destroys the topological protection of the edge states\cite{QSHobs}  regardless of the material qualities of the system. The difficulty of directly probing the spin structure of the edge modes  has stimulated several works studying the behaviour of Quantum Spin Hall edge states when time-reversal symmetry is broken by an external magnetic field\cite{QSHmagnetic,QSHexchange}.

The possibility of driving quantum phase transitions via application of an in-plane magnetic field or magnetic doping has been explored in both two-dimensional (2D)\cite{QSHmagnetic,QSHexchange} and three-dimensional\cite{QSHtoQAH,TiThinFilm,QAHpred} topological insulators.  2D systems in topological phases with broken time-reversal symmetry are Chern insulators, which possess nonzero Chern numbers\cite{Haldane} (the integral of the Berry curvature over the Brillouin Zone).  The bulk topology is manifested at the boundary of the system in the form of one-dimensional modes which move along only one direction at each edge and are protected from backscattering, with the Chern number counting the number of edge modes\cite{BulkBoundary}. The properties of the Chern insulating phase do not depend on time-reversal symmetry and are therefore robust to magnetic disorder, and can be probed by charge transport measurements (e.g. in a four-terminal geometry typically used for measurement of the quantum Hall effect). This state of matter was first proposed in a hexagonal tight-binding model by Haldane\cite{Haldane}. Despite, however, decades of theoretical proposals for its realization in honeycomb materials\cite{grapheneRashba,graphenestaggering,Graphenestrong,BLG,Silicene,Silicene2}, magnetically doped topological insulator thin films\cite{QSHtoQAH,TiThinFilm,QAHpred}, 2D semiconductor systems\cite{QSHexchange,QSHmagnetic,QAHnanopatterned} and optical lattices\cite{OptLatt}, it was only recently observed in a magnetically doped BiTe thin film\cite{QAHobs}. In this work we demonstrate that the Chern insulating state is experimentally accessible in hexagonally patterned $p$-type semiconductor heterostructures, which are quantum simulators of graphene with potentially strong and tunable spin-orbit interactions\cite{Polini}. In absence of a magnetic field, the system is in a topologically insulating state\cite{SushkovCastroNeto}. The Chern insulating state is created by application of an in-plane magnetic field, which is used to break time-reversal symmetry and drive the system through a topological phase transition.



The system we consider is a hexagonally patterned 2D $p$-type semiconductor heterostructure grown along the (311) interface, described by the Hamiltonian
\begin{gather}
H = H_{\text{Lutt.}} + U(x, y) - 2\kappa \mu_B \bm{B} \cdot \bm{S} \ \ ,
\label{hamil2D}
\end{gather}
where $U(x, y)$ is the hexagonal superlattice potential, $-2\kappa \mu_B \bm{B} \cdot \bm{S}$ is the Zeeman coupling due to an in-plane magnetic field $\bm{B} = (B_x, B_y, 0)$, $\bm{S}$ are the spin-$\frac{3}{2}$ operators acting on the $p_{\frac{3}{2}}$ type valence band states, and $H_{\text{Lutt.}}$ is obtained from the 3D Luttinger Hamiltonian for the valence band of bulk zincblende semiconductors\cite{Luttinger} via 2D confinement in an infinite square well in the $z$-direction,
\begin{gather}
H_{\text{Lutt.}} = 
- \frac{ \gamma_2 \langle k_z^2 \rangle S_z^2}{m_e} + (\gamma_1 + \gamma_2( S_z^2 - \frac{5}{4})) \frac{k_x^2 + k_y^2}{2m_e} \nonumber \\
- \frac{ \gamma_2}{m_e} ( \frac{k_+^2 S_-^2 + k_-^2 S_+^2}{4} + \delta \langle k_z^2 \rangle \{S_y, S_z\}) \ \ .
\label{lutt}
\end{gather}
We have taken the projection onto the lowest transverse level via the substitutions $k_z \rightarrow \langle k_z \rangle = 0$, $k_z^2 \rightarrow \langle k_z^2 \rangle = \frac{ \pi^2}{d^2}$ where $d$ is the width of the quantum well, and choose coordinates $x\parallel (0\bar{1}1), y \parallel (\bar{2} 33), z\parallel (311)$, with $k_\pm = k_x \pm i k_y$, \emph{etc}. Here $\gamma_1, \gamma_2$ are the Luttinger parameters\cite{Vurgaftman}, and $\delta$ is a parameter characterizing the cubic anisotropy which is present for $p$-type systems only when the heterostructure is grown along a low-symmetry direction, for a (311) GaAs quantum well \cite{LiYeoh} $\delta \approx 0.1$.

A hexagonal potential of the form
\begin{gather}
U(x, y) = 2W ( \cos \bm { G_1} \cdot\bm{ r} + \cos \bm{ G_2} \cdot\bm{ r} + \cos \bm{ G_3} \cdot\bm{ r})  \ \ ,
\label{lattice}
\end{gather}
with reciprocal lattice vectors $\bm{G_1} = \frac{4 \pi}{\sqrt{3} a} \bm{\hat{y}}, \ \  \bm{G_2} = \frac{4 \pi }{\sqrt{3} a} ( \frac{\sqrt{3}}{2} \bm{ \hat{x}} + \frac{1}{2} \bm{\hat{y}}), \ \  \bm{G_3} = \frac{4 \pi}{\sqrt{3} a} ( - \frac{\sqrt{3}}{2} \bm{ \hat{x}} + \frac{1}{2} \bm{\hat{y}})$, creates minibands with a hexagonal Brillouin Zone. At $\bm{B}= 0$ the dispersion is graphene-like, with Dirac points occurring at the $K$ and $K'$ points at the corners of the Brillouin Zone\cite{DiVincenzo}. At hole density corresponding to filling of two holes per unit cell, the Fermi energy coincides with the lowest Dirac point, which is $ p = 1.4\times 10^{11} \text{cm}^{-2}$ for superlattice spacing $a = 40 \text{nm}$. The momentum corresponding to the Dirac point is $k_\parallel = K = \frac{4\pi}{3a}$, so that for a typical well width $d = 20 \text{nm}$ the ratio $\frac{k_\parallel^2}{\langle k_z^2\rangle} \approx 0.44$ and the dominant spin-dependent interaction is proportional to $- \langle k_z^2 \rangle S_z^2$ and forms quantum states $|S_z = \pm \frac{3}{2}\rangle $ and $|S_z = \pm \frac{1}{2}\rangle $, and the energy splitting between them is $\Delta_{\frac{3}{2} - \frac{1}{2}} = \frac{2 \gamma_2 \langle k_z^2 \rangle}{m_e}$. The $|\pm \frac{3}{2}\rangle$ and $|\pm \frac{1}{2}\rangle$ states have different effective masses, we consider only the $|\pm \frac{3}{2}\rangle$ states, which are lowest in energy, and have mass $m =\frac{m_e}{\gamma_1 + \gamma_2}$. The terms containing $k_+^2 S_-^2, k_-^2 S_+^2$ provide a momentum-dependent coupling between the $|\pm \frac{3}{2} \rangle$ and $|\pm \frac{1}{2}\rangle$ states, and therefore generate the Berry curvature which is necessary to observe non-trivial topological states.

We will describe physics near the $K$ and $K'$ points using a low-energy effective Hamiltonian which acts on a pair of coordinate wavefunctions\cite{SushkovCastroNeto} $|a\rangle, |b\rangle$ and the  spin states $|\pm \frac{3}{2} \rangle$, which we take to be effectively a spin-$\frac{1}{2}$ doublet, $|\frac{3}{2} \rangle = |\uparrow\rangle, |-\frac{3}{2} \rangle = |\downarrow\rangle$, and use the spin-$\frac{1}{2}$ operators $\bm{s}$ so that $s_z|\uparrow \rangle = \frac{1}{2} |\uparrow\rangle, \ \  s_z |\downarrow\rangle = -\frac{1}{2} | \downarrow\rangle$, and the Pauli matrices $\bm{\sigma}$ with $\sigma_z |a\rangle = |a\rangle, \ \ \sigma_z |b\rangle = -|b\rangle$. We choose the basis of coordinate wavefunctions\cite{footnote}
\begin{gather}
|a\rangle = \frac{1}{\sqrt{3}} ( e^{ i\tau_z \bm{K}_1 \cdot \bm{r}} + e^{- \frac{2\pi i}{3}} e^{ i\tau_z\bm{K}_2 \cdot \bm{r}} + e^{ \frac{2 \pi i}{3}} e^{i \tau_z\bm{K}_3 \cdot \bm{r}}) \ \ , \nonumber \\
|b\rangle = \frac{1}{\sqrt{3}} ( e^{ i\tau_z \bm{K}_1 \cdot \bm{r}} + e^{ \frac{2\pi i}{3}} e^{ i \tau_z\bm{K}_2 \cdot \bm{r}} + e^{ -\frac{2\pi i }{3}} e^{i \tau_z\bm{K}_3 \cdot \bm{r}})
\label{basis}
\end{gather}
where $\tau_z = +1$ for the $K$ valley and $-1$ for the $K'$ valley, $\bm{K_1} = K \bm{\hat{x}}, \ \ \bm{K_2} = - \frac{K}{2} \bm{\hat{x}} + \frac{ \sqrt{3} K}{2} \bm{\hat{y}}, \ \ \bm{K_3} = - \frac{ K}{2} \bm{\hat{x}} - \frac{\sqrt{3}K}{2} \bm{\hat{y}}$ are the equivalent momenta corresponding to the $K$ point. The effective Hamiltonian near the Dirac points is given by
\begin{gather}
H = v \tau_z \bm{q} \cdot \bm{\sigma} - 2 \eta \sigma_z s_z - \beta_- \sigma_+ s_+ - \beta_+ \sigma_- s_- - 2\gamma s_z \ \ ,
\label{hamilChern}
\end{gather}
where the Kane-Mele\cite{KaneMele} term $-2\eta \sigma_z s_z$ arises from the spin-orbit terms $k_+^2 S_-^2 +h.c.$ in (\ref{hamil2D}) and leads to a Quantum Spin Hall Insulating state at zero magnetic field\cite{SushkovCastroNeto}, and $\beta_+, \beta_-$ are proportional to the applied in-plane magnetic field. The parameters $\eta, \beta, \gamma$ are given in second order perturbation theory by
\begin{gather}
v = \frac{3 a}{4\pi m} \ \ , \ \ \eta = \frac{ 8W}{9} (\frac{d}{a})^4 \ \ , \nonumber \\
\beta_\pm = \frac{ 4 \kappa \mu_B B_\pm}{3} ( \frac{d}{a})^2 \ \ , \ \ 
\gamma = 3 \kappa \delta \mu_B B_y \ \ .
\end{gather}
These parameters may be renormalized by higher orders of perturbation theory, although accurate band structure calculations show that they are good approximations for the lowest Dirac point.

Noting that the hexagonal potential satisfies $U(x,y) = U(-x,-y)$, the system obeys both time-reversal and inversion symmetry at zero magnetic field, and each band is exactly doubly degenerate, with the band extremum states being simultaneous eigenstates of spin and pseuodspin, $|\sigma_z, s \rangle$. Time reversal symmetry is broken in the magnetic field, leading to a splitting of the originally degenerate bands. At sufficiently large magnetic fields, a topological phase transition is possible if the bulk gap is closed. We may understand the distinct phases of the system treating $\beta_x, \beta_y, \gamma$ as independent parameters. There are four bands, $E_n(\bm{q})$, which at $\bm{q} = 0$ correspond to the states and energies
\begin{gather}
\psi_1 = \cos \frac{\zeta}{2} |a\rangle|  \uparrow \rangle + \sin \frac{\zeta}{2} |b\rangle| \downarrow\rangle \ \ ,\nonumber \\
 \ \ E_1(\bm{q} = 0) = - \eta - \sqrt{ \gamma^2 + 4\beta^2} \  \ ,
 \ \  \tan \zeta = \frac{ 2\beta }{\gamma} \nonumber \\
\psi_2 = - \sin \frac{\zeta}{2} |a\rangle| \uparrow\rangle + \sin \frac{\zeta}{2} |b\rangle |\downarrow\rangle \ \ , \nonumber \\
\ \ E_2(\bm{q} = 0) = - \eta + \sqrt{ \gamma^2 + 4\beta^2} \ \ ,  \nonumber \\
\psi_3 = |b\rangle |\uparrow\rangle \ \ , \ \ E_3(\bm{q} = 0) = \eta - \gamma \nonumber \\
\psi_4 = |a\rangle |\downarrow \rangle \ \ , \ \ E_4 (\bm{q} = 0)= \eta + \gamma \ \ ,
\end{gather}
where $\beta = \sqrt{\beta_x^2 + \beta_y^2}$.

For small values of $\beta, \gamma$, we have $E_2( 0) < E_3(0)$. At a critical set of parameters $\beta^2 + \eta |\gamma| - \eta^2 = 0$, we have $E_2(0) = E_3(0)$ and the system is gapless. This gapless phase forms a boundary between two distinct insulating phases characterized by different edge properties. The evolution of edge modes in the parameter space ($\beta, \gamma$) is illustrated in Fig. 1 (the dispersion of edge modes as a function of the momentum $q_x$ along the edge is shown in solid lines, while the bulk dispersion is shown in dashed lines as a function of $q_x$ for fixed values of $q_y$). For subcritical magnetic fields, $\beta^2 + \eta|\gamma| < \eta^2$, the system is characterized by counterpropagating edge modes in the insulating gap and absence of local current at each edge (Fig. 1a, 1b). These counterpropagating modes are inherited from the topological insulating state\cite{SushkovCastroNeto} at $\beta = \gamma = 0$, but are not protected from backscattering from nonmagnetic disorder\cite{QSHedgeprotected}. In this phase the Chern number is $N = 0$. For parameters close to the critical state (Fig. 1b), the edge mode at the $K'$ valley terminates at the edge of the $E_2$ band (i.e. at the lower edge of the gap), while the edge mode at the $K$ valley is split into two branches, with the left branch lying above the top of the $E_3$ band (i.e. above the gap), while the right branch extends through the gap, terminating slightly above the edge of the left mode. These three branches survive both in the gapless critical state (Fig. 1c) and in the Chern insulating state (Fig. 1d). The edge modes at energy slightly above the band-meeting point in the gapless state are shown above the plots in Fig. 1c, with two clockwise propagating modes at the $K$ valley and a single counterclockwise propagating mode at the $K'$ valley.

The Chern insulating state (Fig. 1d) results from the inversion of the $E_2$ and $E_3$ bands, generating a non-trivial bulk topology. In this state both clockwise moving modes at the $K$ valley extend throughout the gap, while the originally counterclockwise moving mode at the $K'$ valley dips below the bulk dispersion of the $E_4$ band, leading to two counterpropagating modes which exist only at energies close to the top of the insulating gap, disappearing at lower energies. The counterpropagating modes at the $K'$ valley are not shown in the figure above the plots in Fig. 1d. Thus there exists a pair of edge states in the gap which propagate clockwise around the boundary of the system without interruption and carry nonvanishing local current at each edge.

The qualitative change in the behaviour of edge states indicates a topological phase transition from a state with $N = 0$ (Fig. 1a) to a Chern insulator (Fig. 1b). For $\gamma > 0$ (the case shown in Fig. 1b), current flows clockwise around the sample, while the direction is reversed (anticlockwise) for $\gamma < 0$. A pair of co-propagating modes in the insulating gap corresponds a non-trivial Chern number $N = 2$ for clockwise propagation and $N = -2$ for counterclockwise propagation. The phase diagram of the system in the parameter space $\frac{\gamma}{\eta}$, $\frac{ \beta}{\eta}$ is shown in Fig. 2. The $N = 0$ phase occurs for parameters $\beta^2 + \eta \gamma < \eta^2$, and Chern insulating phases with $N = +2 \ (-2)$ exist for $\beta^2 + \eta \gamma > \eta^2$ and $\gamma > 0  \ (<0)$. Gapless phases exist along the boundaries, $\gamma = 0, \beta > \eta$ and $\gamma = \frac{ \eta^2 - \beta^2}{\eta} \ , \beta < \eta$, corresponding to topological semimetallic phases whose properties we will explore in future studies. Unlike many previous proposals for the realization of a Chern insulator\cite{grapheneRashba,QAHnanopatterned,QSHtoQAH,TiThinFilm,QAHpred} our system is fully symmetric under spatial inversion $\bm{r} \rightarrow -\bm{r}$, with the chirality of edge states is determined by the sign of $\gamma$, which in the original model (\ref{hamil2D}) is $\propto B_y$.

\begin{figure}\centering
	\includegraphics[width = 0.4\textwidth]{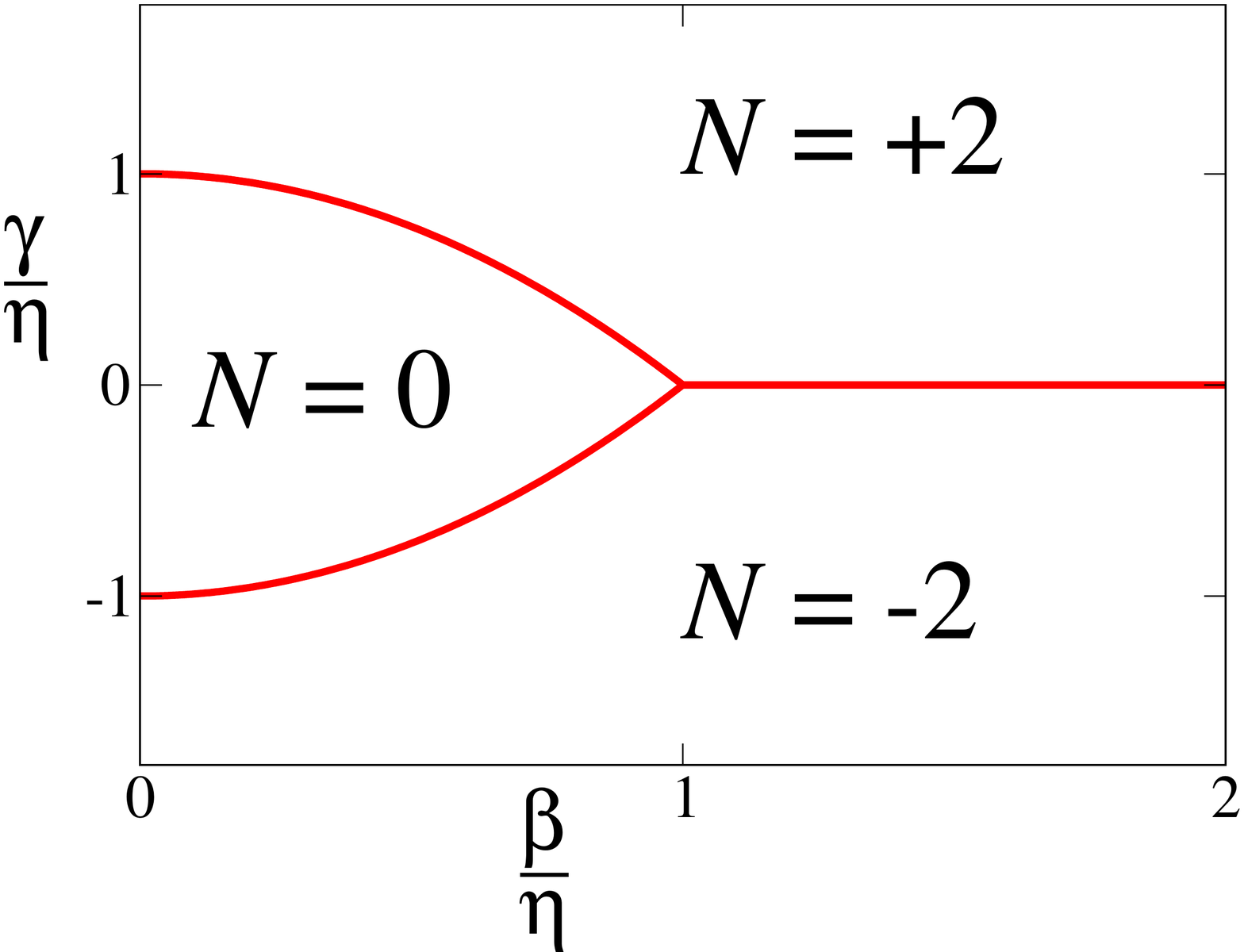}
	\caption{The phase diagram in the parameter space $(\gamma, \beta)$, showing three topologically distinct insulating phases characterized by Chern numbers $N = +2, -2, 0$. The red line indicates gapless phases at the boundary between insulating phases.}
\end{figure}

\begin{figure*}
	\begin{tabular}{cc}
		\includegraphics[width = 0.42\textwidth]{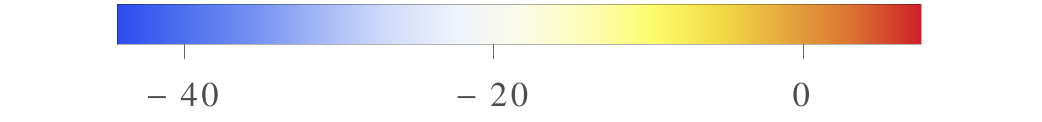} &
		 \includegraphics[width = 0.42\textwidth]{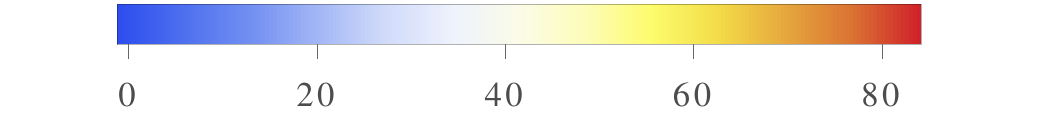} \\
		 \includegraphics[width = 0.42\textwidth]{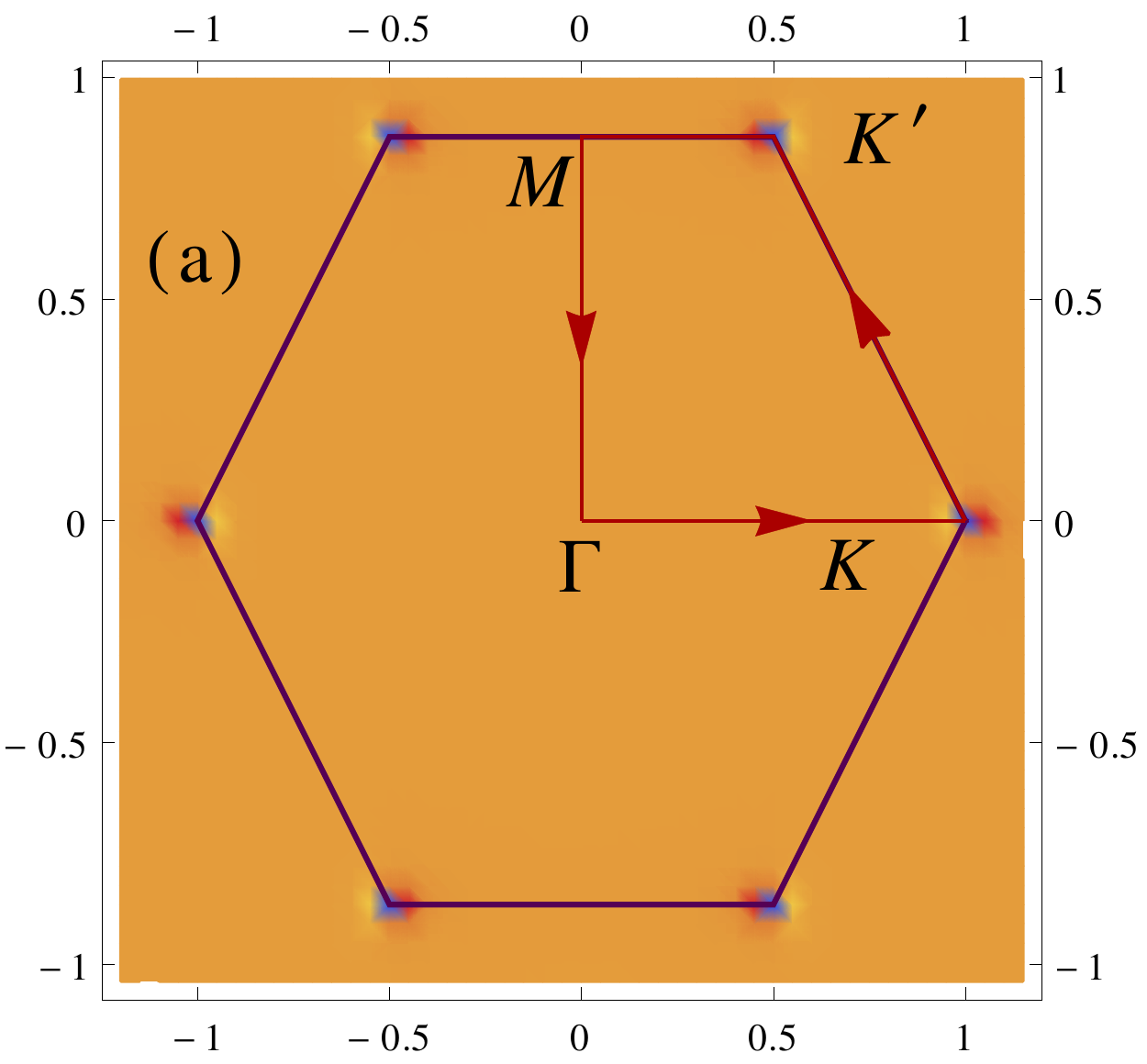} &
		 \includegraphics[width =0.42\textwidth]{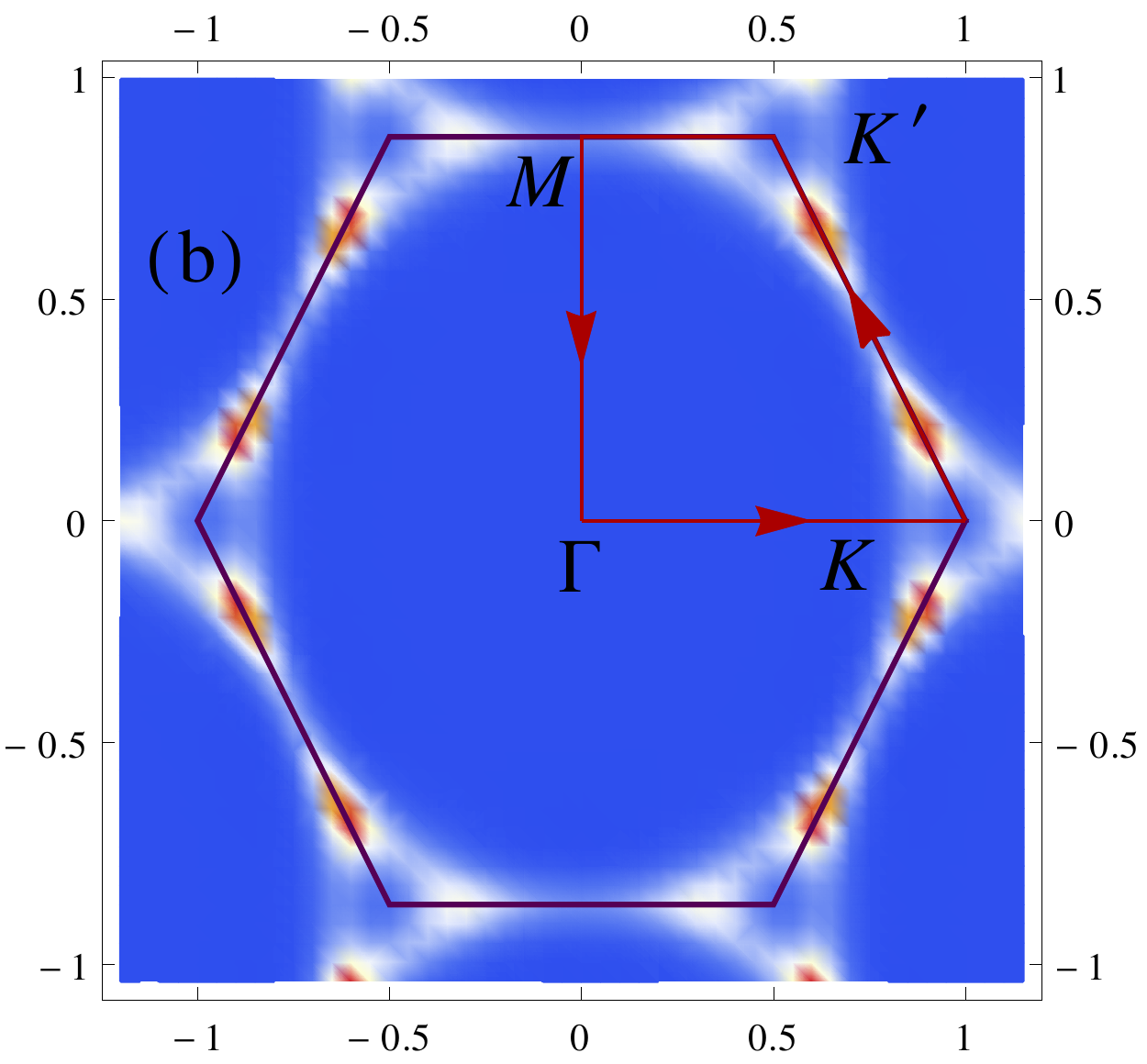} \\ 
		\includegraphics[width = 0.47\textwidth]{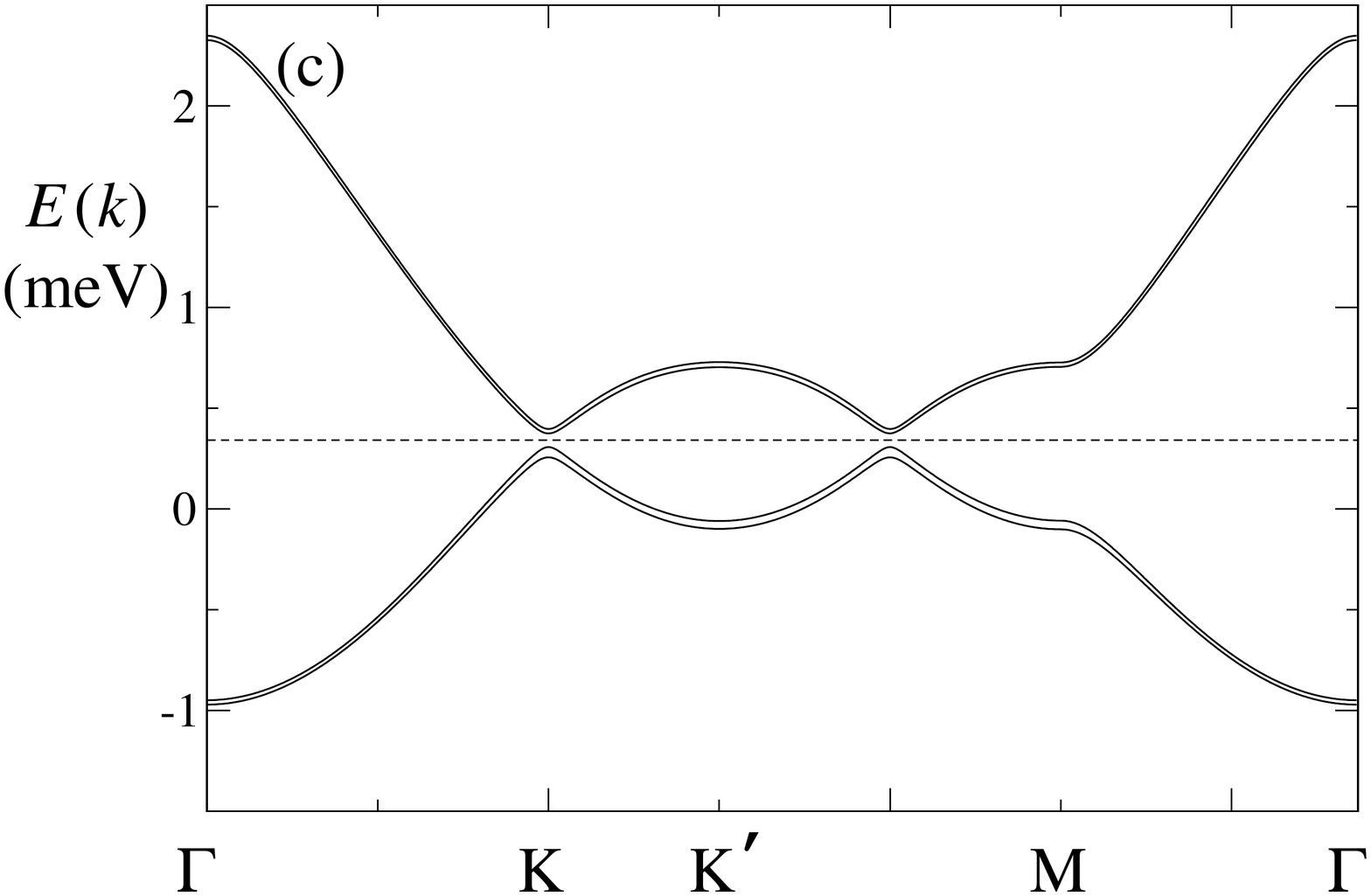} &
		\includegraphics[width = 0.47\textwidth]{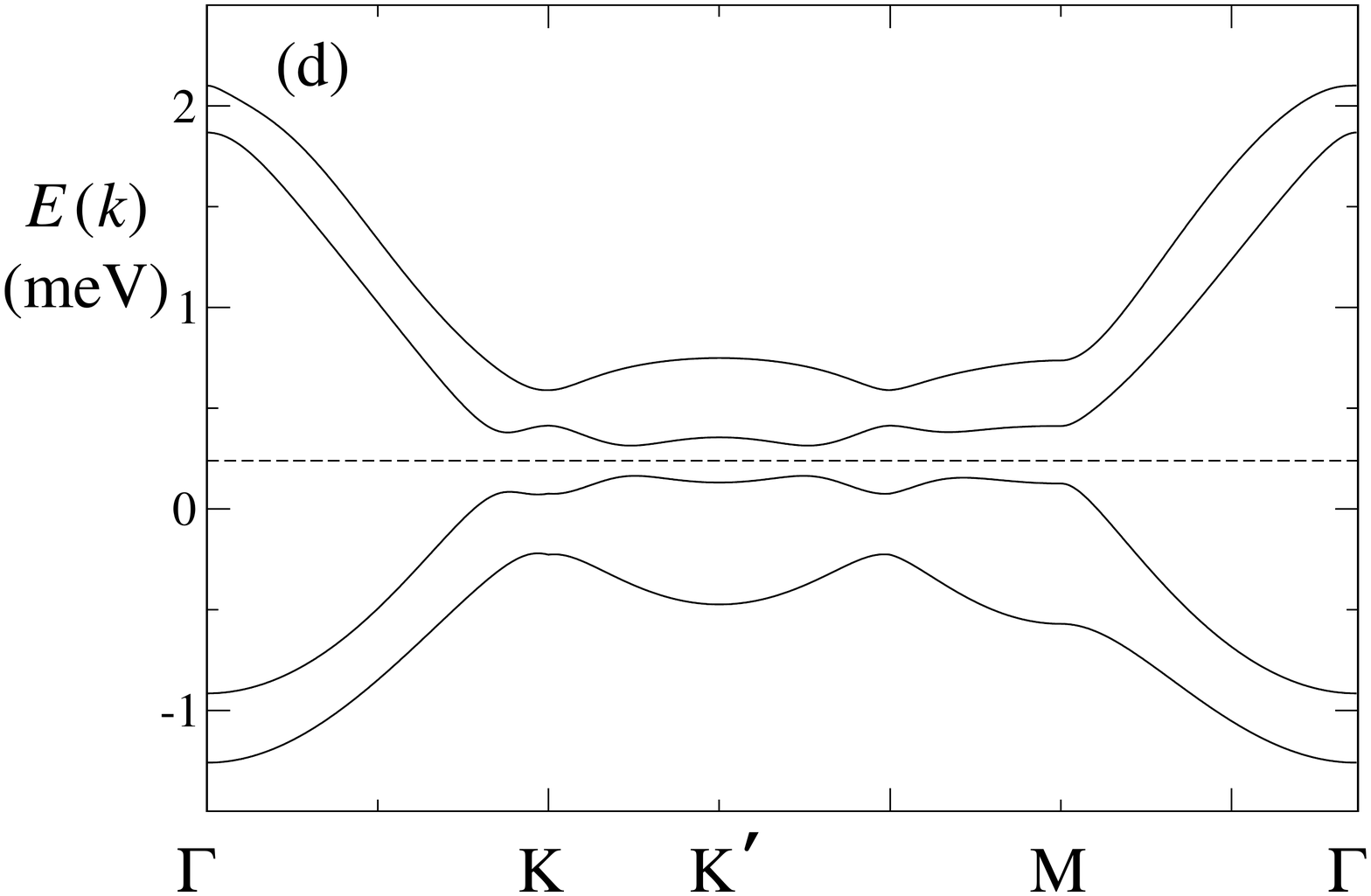}
	\end{tabular}
	\caption{(a),(b): Maps of the Berry curvature summed over the lowest two bands, calculated via numerical solution of the free-electron model with Hamiltonian (\ref{hamil2D}), for the (a) the $N = 0$ phase, with $B_x = 0\text{T}, B_y = 0.5\text{T}$ and (b) the $N = 2$ Chern insulating phase, with  $B_x = 0\text{T}, B_y = 8\text{T}$. The parameters of the system are $a = 40 \text{nm}, d = 20 \text{nm}, W = 1\text{meV}$. The Berry curvature is shown in units where $K = \frac{4\pi}{3a} = 1$. (c), (d): The dispersion relation for the lowest eight bands for the same parameters as the figures above, calculated along the directions indicated by the red line and arrows shown in Panels a, b. The chemical potential is indicated by the dashed line.}
\end{figure*}


In order to realize the Chern insulating phases, we need to have either $\beta \gtrsim \eta$ or $\gamma \gtrsim \eta$. The size of the gap in the Chern insulating phase is $\Delta \approx \gamma = 3\kappa \delta \mu_B B_y = 0.12\text{meV}$ at $B_y = 5\text{T}$ in GaAs. Due to the small size of the gap, it is necessary to account for the effects of disorder in the system. In ultrahigh mobility GaAs hole systems, the mean free path can be $l \approx 3\mu\text{m}$ in the unpatterned device\cite{Jason}, corresponding to an inverse carrier lifetime $\tau^{-1} = 12\mu\text{eV}$. For an artificial lattice created by perforated gates, the most significant disorder in the system is therefore random fluctuations in the potential $U(x,y)$ (\ref{lattice}) due to imperfections of the  gates. To obtain well separated bands we require $W$ to be the same order as the kinetic energy at the $K$,$K'$ points, $W \sim E_0 = \frac{ 8\pi^2}{9m a^2}$ where $m \approx  0.2m_e$ is the 2D hole mass in GaAs. For a superlattice of spacing $a = 40\text{nm} \ll l$ and $d =20\text{nm}$, with $W = 0.5E_0 = 1.0\text{meV}$, the gap in the topological insulating phase is $2\eta = 0.10\text{meV}$, and the Chern insulating transition for $B_x = 0$ occurs at $B_y = 1.4\text{T}$. In order to observe a gap of $\Delta \approx 0.1 \text{meV}$ with these parameters, it is necessary to suppress fluctuations in the potential below 10\% of the average value.

The existence of chiral edge modes is deeply connected to the topology of the 2D Bloch states and to the transverse conductivity $\sigma_{xy}$\cite{Laughlin}. Let us suppose that an electric field is applied along the $y$-direction, corresponding to a positive voltage difference $eV_y$, which changes the chemical potential of right and left movers along the edges of the system. For $N > 0$ (clockwise propagation), this leads to a current $J_x > 0$ due to the  excess of right-moving holes along the top edge. The conductivity is given by $\sigma_{xy} = \frac{ e^2 N}{h}$ where $N$ is the Chern number, and is exactly quantized (at zero temperature and in the absence of inelastic scattering) due to the absence of backscattering for chiral 1D modes. The Kubo formula applied to the transverse conductivity can be expressed in terms of the Berry curvature of filled bands\cite{TKNN},
\begin{gather}
\sigma_{xy} = \frac{e^2}{h} \sum_{n, E_n < \mu} {\frac{1}{2\pi}\int{ F_{n,xy} (\bm{k}) dk_x dk_y} } \ \ , 
\end{gather}
which implies the relation between the number and direction of edge states and the topology of bulk states, $N = \frac{1}{2\pi}\sum_n{ \int{ F_{n, xy}(\bm{k}) dk_x dk_y}}$. The bulk-boundary correspondence, which relates the exact quantization of electronic transport coefficients and the existence of robust edge states has been rigorously extended to topological systems in arbitrary dimensions with local symmetries\cite{BulkBoundary}. In order to illustrate this correspondence, we have calculated the Berry curvature $F_{n,xy}$ from numerical solution of the nearly-free-electron model with the Hamiltonian (\ref{hamil2D}). Maps of the Berry curvature summed over the lowest two bands are shown in Fig. 3a,b for $B_x = 0, B_y = 0.5\text{T}$ (the $N = 0$ phase) and $B_x = 0, B_y =   8\text{T}$ (the $N = 2$ Chern insulating phase). The dispersion is shown in panels c,d of the same figure underneath the panels corresponding to the same parameters and along the paths indicated by the arrows in Fig. 3c,d. The extreme values of the Berry curvature for the two values of $B_y$ differ significantly, so we have chosen different color schemes for the two plots. For  $B_y = 0.5\text{T}$, integration of the Berry curvature over the Brillouin Zone gives $\int{ F_{xy} dk_x dk_y} = -2\pi$ and $+2\pi$ for the lowest and second lowest bands respectively. The orange regions correspond to areas in the Brillouin Zone for which the Berry curvature is close to zero,  and red and blue spots correspond to the maxima and minima of the Berry curvature respectively. In the Chern insulating state ($B_y = 8\text{T}$), the Berry curvature integrated over the lowest band is $-2\pi$, but in the second lowest band is equal to $+6\pi$, which is consistent with Chern number $N = 2$ as calculated explicitly from the edge dispersion. The region in which the total Berry curvature is close to zero is shown in blue, while the red spots indicate points in the Brillouin Zone in which Berry curvature is peaked. This occurs at anticrossings in the dispersion in which the splitting between bands is small and the wavefunction is quickly varying (see Fig. 4d). Upon switching the sign of the magnetic field to $B_y = -8\text{T}$, the dispersion (Fig. 3d) remains the same, but the Berry curvature in the lowest and second lowest bands are $+2\pi$ and $-6\pi$ respectively, consistent with Chern number $N = -2$. These numerical results demonstrate the bulk-boundary correspondence for topological systems\cite{BulkBoundary}.

Our results illustrate the versatility of hexagonally patterned 2D hole systems due to the highly tunable spin-orbit interaction. We have calculated the topological phase diagram of the system in the presence of an in-plane magnetic field and cubic anisotropy, and discussed the properties of the non-trivial insulating phases. The magnetic field drives transitions between insulating phases classified by different Chern numbers $N = 0$ at low magnetic fields and $N = +2, -2$ at high magnetic fields. The Chern insulating phases are characterized by co-propagating clockwise ($N = 2$) or anticlockwise ($N = -2$) edge modes, Hall conductivities $\sigma_{xy} = +\frac{2e^2}{h}, -\frac{2e^2}{h}$ without a perpendicular magnetic field, with the direction edge currents controlled by the orientation of the in-plane field.
The presence of topological phases which can currently be experimentally realized in hexagonally patterned $p$-type semiconductor heterostructures in both the presence and absence of time-reversal symmetry suggests the exciting possibilities of further expanding the topological phase diagram in the interacting scenario, e.g. to realize fractional topological insulating\cite{FracTI} or topologically superconducting phases\cite{TopSup}.

\appendix

\section{Calculation of edge states for a hard-wall potential}

The edge states shown in Fig. 1 were evaluated for a hard-wall potential along the $x$-direction, $V(y > 0) = 0 \ \ , \ \ V(y < 0) = \infty$. The momentum along the edge $q_x$ is a good quantum number, and the wavefunction has the form
\begin{gather}
\psi = C_+ e^{i q_x x - \kappa_+ y} \psi_+ + C_- e^{i q_x x - \kappa_- y} \psi_-
\end{gather}
where $\kappa_+, \kappa_-$ are generally complex, with the wavefunction being exponentially localized at the edge when the hard-wall condition $\psi(x, y= 0)$ is satisfied and the energy lies in a range where $\text{Re} \kappa_\pm > 0$. The states $\psi_\pm$ are eigenvectors of the Hamiltonian (6) with energy $E$ and complex $q_y = i \kappa_\pm$ satisfying 
\begin{gather}
\kappa_\pm =
\nonumber \\
( v^2 q_x^2  - E^2 - \gamma^2 - \eta^2 \pm 2i \sqrt{ -\gamma^2 E^2 - b^2 ((E - \eta)^2 - \gamma^2) })^{\frac{1}{2} } \ \ .
\end{gather}
Taking into account the real-space structure of the basis states (5), the boundary condition implies that
\begin{gather}
( \langle a, s_z| \psi_+ \rangle + \langle b, s_z| \psi_+ \rangle ) C_+ + ( \langle a, s_z | \psi_- \rangle + \langle b, s_z | \psi_-\rangle ) C_-
\nonumber \\
= 0
\end{gather}
for $s_z = \pm\frac{1}{2}$. Solutions $C_+, C_-$ exist when
\begin{gather}
f(q_x, E) =
\tau_z \gamma E(E - \eta + v \tau_z q_x)(\kappa_+ - \kappa_-) + 
\nonumber\\
i \sqrt{- \gamma^2 E^2 - \beta^2((E - \eta)^2 - \gamma^2)}\times 
\nonumber \\ ( (E - \eta + v \tau_z q_x)(E - \eta + v \tau_z q_x) \nonumber \\
- \kappa_+ \kappa_- 
+ \tau_z \gamma (\kappa_+ + \kappa_-) - \gamma^2 ) = 0 \ \ .
\label{dispedge}
\end{gather}
Regarding $\kappa_\pm$ as a function of $q_x, E$, the equation $f(q_x, E) = 0 $ provides the implicit dispersion relation of the one-dimensional edge states. There are two solutions for $\tau_z = +1$ which close the gap and propagate to the left, $\frac{dE}{dq_x} < 0$, in addition to a pair of solutions for $\tau_z= -1$ corresponding to counter-propagating modes which disappear when the energy becomes sufficiently close to the lower ($E_3(\bm{q})$) band. The energy can be chosen to intersect only the co-propagating modes.

The edge states in Fig. 1 correspond to modes which propagate along the bottom edge of the sample. We may calculate the edge states for the top edge in the same way, taking the edge potential $V(y < L_y = 0), V(y > L_y) = \infty$ where $L_y= \frac{ \sqrt{3} na}{2}, n = 1, 2, \dots$. In order for these states to be exponentially localized, we require $q_y = -i \kappa_\pm$ with $\text{Re} \kappa_\pm > 0$. This implies that the condition for the existence of edge states on the top edge is identical to Eq. (\ref{dispedge}) after substitution $\kappa_\pm \rightarrow - \kappa_\pm$. Equivalently, we may perform a substitution $q_x \rightarrow - q_x, \tau_z \rightarrow - \tau_z$ to obtain the dispersion of edge states along the top of the system. This implies that the 1D modes at the top of the system will exist at opposite valleys and propagate in the opposite direction to the modes along the bottom of the system. The modes along the left and right edges, corresponding to hard-wall potentials $V_L(x < 0) = +\infty, V_L(x > 0) = 0$ and $V_R(x > L_x) = +\infty, V_R(x < L_x ) = 0$ with $L_x = \frac{ n a}{2}, n=1, 2, \dots$, may be calculated in a similar fashion using the boundary conditions $\psi(x = 0) = \psi(x = L_x) = 0$. The edge states have identical structure to those along the top and bottom, with the left mode propagating along the $+y$ direction and the right mode propagating along the $-y$ direction.


\begin{thebibliography}{99}
\bibitem{footnote}
Note that, in some previous literature the basis wavefunctions $|a\rangle,|b\rangle$ are chosen differently. We choose a helical basis, so that the linear-in-momentum term Hamiltonian is equal to $v\tau_z\bm{q} \cdot \bm{\sigma}$ as in Eq. (\ref{hamilChern}).
\end{thebibliography}
\end{document}